# Robot-Initiated Social Control of Sedentary Behavior: Comparing the Impact of Relationship- and Target-Focused Strategies

Jiaxin Xu
*Human-Technology Interaction Group*
*Eindhoven University of Technology*
Eindhoven, The Netherlands
j.xu2@tue.nl

Sterre Anna Mariam van der Horst
*Human-Technology Interaction Group*
*Eindhoven University of Technology*
Eindhoven, The Netherlands
vanderhorst.sterre@gmail.com

Chao Zhang
*Human-Technology Interaction Group*
*Eindhoven University of Technology*
Eindhoven, The Netherlands
c. zhang.5@tue.nl

Raymond H. Cuijpers
*Human-Technology Interaction Group*
*Eindhoven University of Technology*
Eindhoven, The Netherlands
r.h.cuijpers@tue.nl

Wijnand A. IJsselsteijn
*Human-Technology Interaction Group*
*Eindhoven University of Technology*
Eindhoven, The Netherlands
w.a.ijsselsteijn@tue.nl

*Abstract*—To design social robots to effectively promote health behavior change, it is essential to understand how people respond to various health communication strategies employed by these robots. This study examines the effectiveness of two types of social control strategies from a social robot—relationship-focused strategies (emphasizing relational consequences) and target-focused strategies (emphasizing health consequences)—in encouraging people to reduce sedentary behavior. A two-session lab experiment was conducted ($n$ = 135), where participants first played a game with a robot, followed by the robot persuading them to stand up and move using one of the strategies. Half of the participants joined a second session to have a repeated interaction with the robot. Results showed that relationship-focused strategies motivated participants to stay active longer. Repeated sessions did not strengthen participants' relationship with the robot, but those who felt more attached to the robot responded more actively to the target-focused strategies. These findings offer valuable insights for designing persuasive strategies for social robots in health communication contexts.

## I. INTRODUCTION

Unhealthy lifestyle behaviors remain a leading risk factor in the global rise of chronic diseases, making it increasingly urgent to empower individuals to adopt healthier behaviors. As social robots become increasingly integrated into daily life, the human-robot interaction (HRI) community has increasingly focused on their potential as persuasive agents for promoting health behavior change. This focus has inspired various applications, including robots designed to encourage physical activity [22, 59], provide dietary advice [2, 11], assist with weight management [35] and support medication adherence [6, 27].

While these studies provide preliminary evidence that social robots can positively influence health behaviors, relatively few have specifically examined the effectiveness of different health communication strategies employed by these robots. Health communication strategies vary widely, and poorly designed strategies—such as those perceived as threatening freedom—can provoke psychological reactance, diminish motivation, and even lead to counterproductive behaviors [42, 66]. Therefore, designers must carefully craft strategies to ensure they are both effective in fostering compliance and socially appropriate. Moreover, the task of designing persuasive communication for social robots is further complicated by these robots' unique qualities. Unlike traditional persuasive platforms (e.g., browser interfaces or virtual agents), these robots' physical presence and multimodal social behaviors make them particularly adept at evoking emotions and fostering socioemotional relationships with humans [25, 58]. As seen in various applications, social robots increasingly transcend purely instrumental assistants, entering people's personal spaces as playmates, companions, or even close confidants [60]. Given this unique relational evocativeness of social robots, it is imperative to broaden the scope of persuasive communication design by incorporating relational dynamics as an important consideration.

Messages such as "Staying hydrated keeps you energized" or "Exercise is good for your health" exemplify common health communication strategies employed by today's social robots. According to social control theory, these strategies are classified as **"target-focused"** as their primary goal is to inform people about the health consequences of specific behaviors [43]. Their widespread adoption likely stems from the assumption that people are more inclined to accept advice from someone perceived as having expertise or authority, capable of offering useful information [27]. However, another persuasion mechanism remains underexplored in HRI contexts. In interpersonal relationships, people often leverage relational implications to influence others—for example, by expressing "I'll like you more if you don't smoke" or "I'm not sure I can stay with you if you keep smoking." These **"relationship-focused"** strategies have been shown to be particularly effective in promoting behavior change, often outperforming those target-focused ones [17, 18]. This study aims to expand our understanding of how people respond to these different types of social control strategies employed by social robots.

Importantly, persuasion research in interpersonal contexts has consistently shown that the strength of a relationship significantly impacts the effectiveness of persuasion [42, 55]. Social control strategies initiated by close partners tend to be more effective than those delivered by strangers or acquaintances [13, 19, 51]. It is therefore reasonable to assume that the persuasiveness of a robot may similarly depend on the strength of the relationship established between people and the robot. To explore this possibility, this study employed a two-session design aimed at investigating whether repeated interactions could strengthen the human-robot relationship and whether the strength of this relationship moderates the effectiveness of the two contrasting strategies.

## II. BACKGROUND AND RELATED WORK

### A. Health-Related Social Control

People always attempt to influence the health behaviors of others—whether it is a spouse encouraging their partner to stop drinking, a child urging a parent to quit smoking, or a friend motivating someone to exercise. Social psychologists refer to these attempts as health-related social control, recognizing it as a key mechanism through which personal relationships promote health outcomes [44, 69]. The intended impact of social control requires carefully chosen communication strategies [42]. Traditionally, social control strategies are categorized into two broad types based on their emotional valence—positive and negative strategies. Positive strategies, such as providing useful health information, are generally more effective than negative ones like pressuring or criticizing [14, 66]. Despite this foundational understanding, social control researchers have called for more nuanced investigations to better distinguish these strategies and identify the conditions under which they are most effective [9, 68].

A more refined distinction has emerged between relationship-focused and target-focused strategies. In their exploration of common persuasion patterns in interpersonal interactions, Bisanz et al. [5, 62] found that people frequently appeal to relational implications or obligations to influence others, using utterances like "A good friend would do this for me". Equally common are target-focused strategies which emphasize the recipient's self-interest, with utterances such as "This will be good for you." In social control literature, relationship-focused strategies have been found to be more effective than target-focused ones. For example, leveraging relational incentives ("I'd be happier if you ate healthier") is more successful in promoting healthier eating, exercise, and smoking cessation than providing health information. Similarly, strategies that highlight negative relational consequences (e.g., "I'm not sure I can stay with you if you keep smoking") are more effective than offering health facts alone [17, 18].

One of the most critical contextual factors influencing the effectiveness of these strategies is the relationship strength between the social control initiator and the recipient [55]. Persuasion research consistently shows that people are more willing to comply with requests from close partners or friends than from strangers or acquaintances [13, 19, 51], suggesting that relationship intimacy significantly boosts compliance [24, 40, 61]. In a previous study examining various predictors of receptiveness to advice, relationship closeness emerged as the strongest predictor [23]. The psychological mechanism behind this phenomenon is explained by the concept of "transformation of motivation" in interdependence theory [34]. This theory suggests that people in close relationships tend to view healthy actions as meaningful not only for their own health benefits but also as meaningful for their relationship or their partner [73]. When being persuaded, people's motivation shifts from self-centered goals to those that are more pro-relationship, leading them to change their behaviors more readily [43].

### B. Health Communication Strategy in HRI

Research on social robots for health behavior change has explored various communication strategies, yet the findings remain fragmented and mixed. For instance, some studies have shown that robots providing health information [49] or leveraging scientific knowledge [39] can positively encourage behavior change. However, other studies have found that robots relying solely on expertise do not significantly influence people. This may be because robots serving as information sources align too closely with existing expectations, leading to desensitization toward such strategies [71].

A recent study by Xu et al. [72] explored whether a robot using relationship-focused strategies (e.g., "I will like you more if you can be more physically active") could better influence people's intentions to change sedentary behaviors compared to a robot using target-focused strategies (e.g., "Doing some simple exercises will help to relax your body"). Surprisingly, the results indicated that relationship-focused strategies were less effective in triggering intentions to change, were viewed as less appropriate, and caused stronger psychological reactance. However, a notable limitation of the study was that participants only watched videos of the robot. This limitation leaves open the question of how these strategies might influence humans during real human-robot interactions.

Outside of health contexts, some HRI studies suggest that relationship-focused strategies may increase a robot's persuasiveness. In [63, 64], participants played a game with a robot in which they initially made their own decisions, while the robot employed various strategies to persuade them to change these decisions. The results showed that people were more likely to comply with the robot when it said, "It would make me happy if you take my advice," compared to when it employed more logical arguments This finding hints at the possibility that people's behaviors can be influenced by some relationship-oriented motivations, such as a desire to please the robot or maintain a relationship with it.

### C. Human-Robot Relationship

Humans tend to respond to machines socially when those machines display social cues [54]. This response stems from humans' innate tendency to anthropomorphize [74]. As machines become more human-like and exhibit sophisticated social behaviors—as is the case with social robots—they are more likely to evoke deeper emotional responses in humans [16, 25]. As research has shown, when robots engage in self-disclosure, people feel a sense of closeness [31, 50]; when robots exhibit empathic behaviors, they may foster a sense of intimacy [4]; and when robots offer comfort during stressful moments, humans may begin to form emotional bonds with them [70].

These findings offer preliminary evidence that humans can form some kinds of relationships with robots. However, the very notion of "human-robot relationships" remains contested. Critics argue that, since robots are ultimately controlled by humans and lack true emotions of their own, the notion of a relationship—typically built on mutual care and dependence—is either conceptually flawed [21] or deceptive [65]. In response, some HRI researchers suggest that what we call "human-robot relationships" might be more accurately described as one-sided bonds, where people emotionally invest in robots without expecting reciprocity [16, 46]. In this study, we adopt this perspective, defining human-robot relationships as unidirectional emotional perceptions initiated by the human.

Nearly all relationships, whether between humans or humans and non-humans, require time to develop. By nature, relationships are temporal constructs that evolve, either strengthening as familiarity grows or weakening as people drift apart [30]. In human-human interactions, relationship building is seen as progressing through multiple stages, starting from being strangers and gradually deepening into something more meaningful [41]. Similarly, in human-nonhuman interactions, such as those with pets or objects, relationship building requires sufficient and repeated interactions over time [47]. Recognizing the importance of time in relationship building, this study implements repeated interaction sessions, aiming to foster a more robust human-robot relationship.

### D. The Current Study

This study aimed to explore how people's health behaviors are influenced by a robot's different communication strategies. Sedentary behavior was chosen as the target health behavior, as it is a growing global health issue that people from diverse sociodemographic backgrounds can easily recognize. Grounded in social control theory, the study examined the effects of two distinct strategies: relationship-focused strategies which emphasize the relational consequences, and target-focused strategies, which emphasize the health consequences. A lab experiment was conducted in which participants played a seated game with a robot, and afterward, the robot employed different strategies to persuade them to stand up and walk around during a 5-minute break. The first research question (**RQ1**) investigated which of the two strategies would be more effective in promoting health behaviors.

To allow the strategies to fully take effect, the study sought to create opportunities for participants to build a stronger relationship with the robot. A two-session design was implemented, where participants interacted with the same robot twice, with a one-week interval between sessions. The second research question (**RQ2**) examined how repeated sessions might influence the strength of the human-robot relationship. Furthermore, the third research question (**RQ3**) explored how the strength of the relationship might moderate the effectiveness of the robot's social control strategies. Given the exploratory nature of this study, we did not formulate specific hypotheses but rather allowed the data to guide our understanding.

### III. METHOD

This study was approved by the ethics review board at the Human-Technology Interaction group of Eindhoven University of Technology. All participants provided informed consent and were informed about the video recordings during the experiment. Each participant received compensation of 6 euros per session for their participation.

### A. Participants, Design, and Sample Size

This study adopted a two-session mixed-subjects design, with strategy (relationship-focused vs. target-focused) as a between-subjects factor, and session (session 1 vs. session 2) as a within-subjects factor. The sample size was determined by an a-priori power analysis through the superpower package [38] in *R*, based on an estimated raw mean difference of 0.5 minutes in participants' active time, with an estimated standard deviation of 1.25 minutes. According to the analysis for a mixed ANOVA, 70 participants per condition would provide 93.1% power to detect the interaction effect between strategy and session at an alpha level of 0.05. However, recruiting 140 participants for a lab experiment posed significant time and resource constraints. To address this, we opted for a sequential testing approach, splitting data collection into two rounds, with 70 participants per round. An interim analysis was conducted after the first round to determine whether to proceed with further data collection. To mitigate the risk of inflating the Type 1 error rate associated with sequential testing [37], we applied the Pocock correction [56], adjusting the alpha level to 0.0294 for all data analysis. An overview of the study design is in Table I.

In the 1$^{st}$ round of data collection, 69 participants completed the experiment. Three participants were excluded due to technical issues. The remaining 66 were randomly assigned to the relationship-focused (19 Female, 12 Male, 2 not-specified gender; $M_{age}$ = 24.30, $SD_{age}$ = 4.14) and target-focused condition (22 Female, 11 Male; $M_{age}$ = 25.60, $SD_{age}$ = 7.98). The interim analysis revealed no significant increase in relationship strength over sessions, nor a significant interaction effect between strategy and session on active time. However, the data suggested the possibility that strategy might have an effect. Given these findings, and to reduce experimental costs, we proceeded with the 2$^{nd}$ round of data collection but omitted the second session. Another 71 participants were then recruited, with two excluded due to technical issues. Consequently, a total of 135 were included in the final analysis, with 68 in the relationship-focused (37 Female, 27 Male, 4 non-specified gender; $M_{age}$ = 25.00, $SD_{age}$ = 4.36), and 67 in the target-focused condition (42 Female, 25 Male; $M_{age}$ = 26.40, $SD_{age}$ = 7.35).

TABLE I.  OVERVIEW OF THE EXPERIMENTAL DESIGN.

| | *n* | Session 1 | Session 2 |
|---|---|---|---|
| Data collection round 1 | 33 | Relationship-focused | Relationship-focused |
| | 33 | Target-focused | Target-focused |
| **Interim data analysis** | | | |
| Data collection round 2 | 35 | Relationship-focused | - |
| | 34 | Target-focused | - |
| **Final data analysis** | | | |

### B. The Social Robot

The study used the Misty II robot (Misty Robotics, USA) [52]. This desktop-size robot (36 cm tall, 20 cm wide, and 25 cm deep) is equipped with 3-watt stereo speakers, which enable voice interactions. Its 480 × 272 screen allows for the expression of various emotions. Its overall appearance resembles a small baby, with soft contours, a large head, and big eyes (Fig. 1). To align with its appearance, we generated a child-like voice for Misty using the voice generator Narakeet [53], with a gender-neutral, child-like voice, to make the robot more likable.

*C. Task and Manipulation*

The experiment was conducted using the Wizard-of-Oz approach [15], meaning that all of the robot's behaviors were scripted and secretly controlled by an experimenter. The experiment started with a game interaction, after which the robot initiated social control over people's sitting behaviors.

*Game Interaction.* In an effort to support human-robot relationship building, we developed a game task named "Three Questions", inspired by the classic "Twenty Question Game [32]." Our game included eight rounds. In each round, participants were presented with a list of objects related to a specific theme (e.g., animals) and asked to select one on a laptop (Fig. 2). Misty then attempted to guess their choice by asking three questions. Throughout the game, Misty alternated between correct and incorrect guesses. It succeeded in round 1 and failed in round 2 to capture participants' interest. In rounds 3, 4, and 5, Misty guessed correctly and expressed happiness. In round 6, it failed, showing disappointment and asking for emotional comfort, and it succeeded again in the last two rounds. This structure was designed to create a more realistic interaction with a mix of successes and failures while incorporating contextually appropriate social behaviors to foster relationship building (The full interaction scripts are available in the supplementary materials.)

*Social Control Strategy.* The game was intentionally designed as a seated activity, requiring participants to remain seated in a sofa (Fig. 2, left) for approximately 10 minutes. This game duration was chosen to avoid boredom while still providing a valid context for the robot to exert social control over participants' sitting behavior. Specifically, after the game, Misty told participants that it needed 5 minutes to prepare for the questionnaire and encouraged them to stand and walk around during the break using one of the two types of strategies.

Following theoretical definitions [5, 62], the relationship-focused strategies were designed to emphasize the relational consequences of taking an active break. For example, Misty expressed, "It would make me happy if you could stand up and walk around," accompanied by a slightly concerned facial expression to align with its verbal message (Fig. 1, left). In contrast, the target-focused strategies emphasized health benefits through rational evidence, such as, "Research shows that taking active breaks can significantly improve your physical health." For these strategies, Misty maintained a neutral facial expression (Fig. 1, right). All strategies adhered to a consistent structure: starting with a highlight on the situation ("We've been sitting for quite a while now"), leading to the core message, and ending with an explicit call to action ("Stand up and walk around"). In doing so, we aimed to minimize potential confounding factors in the phrasing and ensure that all strategies were of similar length. To avoid excessive repetition, two variations of each strategy type were designed for the two sessions.

*D. Measures*

*1) Behavioral Reaction.* All participants' behaviors during the break were recorded on video. The total observation period was 307s, from when the robot said "stand up" until the 5-minute countdown ended (Fig. 3). A behavioral coding method was applied to analyze these recordings, yielding three measures to assess participants' reactions: a) *Active time*: the total duration participants spent being active during the break. This was calculated by subtracting the time spent sitting from the total 307 seconds (Fig. 3); b) *Standing up:* a binary measure of whether participants stood up after the robot's persuasion (coded as 1 if they stood up and 0 if they did not)*; c) Staying active throughout:* a binary measure indicating whether participants remained active for the entire break duration (coded as 1 if their active time $\geq$ 300s and 0 if it was < 300s).

*2) Human-Robot Relationship.* Four measures were used to assess participants' perceived relationship with the robot. *Attachment* was measured using seven items (e.g. "I felt emotionally connected to Misty") adapted from [29], rated on a 7-point scale (1 = "strongly disagree", 7 = "strongly agree"). The internal consistency was good (Cronbach's $\alpha$ = 0.91). *Closeness* was measured using five items (e.g. "Misty and I are a good match.") adapted from [67]. The internal consistency was good (Cronbach's $\alpha$ = 0.89). *Companionship* was measured by nine 10-point semantic differentials (e.g., 1 = "not loving", 10 = "loving") adapted from [57], also with good internal consistency (Cronbach's $\alpha$ = 0.92). We also developed a new measure *Relationship Stage,* using a sliding scale from 0 ("complete stranger") to 100 ("close friend"), inspired by [3].

*3) Cognitive Evaluation of Strategy.* Two additional constructs were applied to measure the cognitive evaluation of the strategies. *Perceived threat to freedom* was measured with four 7-point items (e.g., "The robot tried to manipulate me") adapted from [20]. *Social appropriateness* was measured using four 5-point scales (e.g., 1 = "socially appropriate"; 5 = "socially inappropriate"), adapted from [8]. The internal reliability of both scales was good (Cronbach's $\alpha$ = 0.82 and 0.85 respectively). (All measurement items are in the supplementary materials).

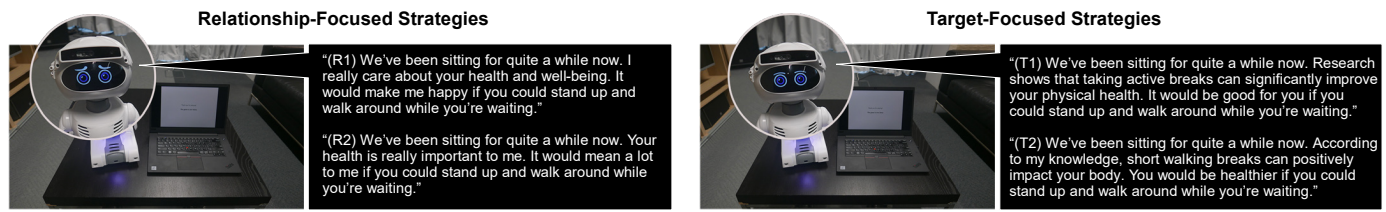

Fig. 1. Design of the relationship-focused strategies (left) and the target-focused strategies (right) for the social robot Misty.

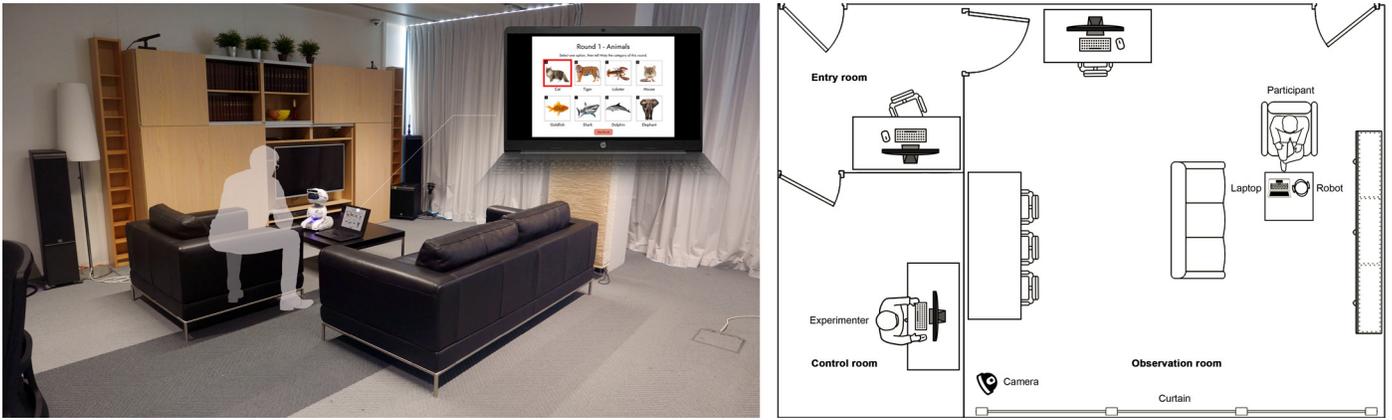

Fig. 2. The experiment setting (left) and lab layout (right).

E. *Setting and Procedure*

The lab consisted of three separate rooms. The entry room contained a desk, a chair, and a computer. Adjacent to this was the observation room, designed to resemble a living room, with a sofa, a laptop, and Misty placed on a tea table. The room measured approximately 30 m$^2$, providing sufficient space for participants to move around. Curtains were drawn to minimize potential confounding influences from outside. The control room, where the experimenter remotely operated Misty, was equipped with monitor equipment (Fig. 2).

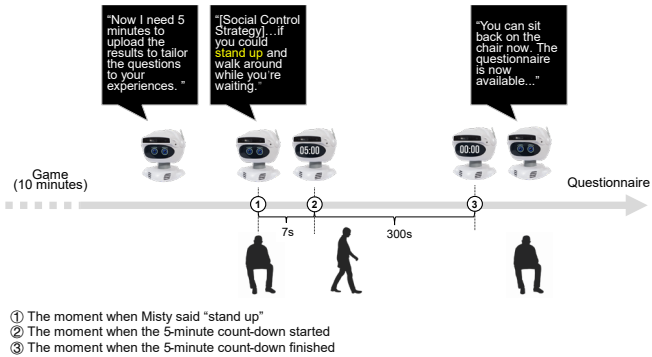

① The moment when Misty said "stand up"
② The moment when the 5-minute count-down started
③ The moment when the 5-minute count-down finished

Fig. 3. Procedure and timeline. The observation period for behavioral reactions lasted a total of 307 seconds, starting when the robot explicitly said "stand up" and ending when the 5-minute countdown timer finished.

*Session 1.* Upon arrival, participants were led to believe that the experiment focused on assessing their physiological data during a gaming interaction with a robot. This deception was designed to prevent participants from modifying their behavior due to the Hawthorne effect [1]. Participants began by signing the informed consent form, after which they were given a wrist-worn accelerometer. The experimenter explained the game rule, and participants were then asked to leave their phones in the entry room and were then escorted to the observation room. Participants played the game with Misty for approximately 10 minutes. After the game, Misty told them that it needed five minutes to prepare a questionnaire and persuaded them to stand up and walk around by using one of the social control strategies. During this break, a 5-minute countdown timer was displayed on Misty's screen. After the break, Misty instructed participants to complete a relationship questionnaire on the laptop. Once finished, Misty bid them farewell.

*Session 2.* Upon arrival, participants received a brief recap of the game rules and were fitted with the accelerometer again. At the beginning of this session, Misty acted as though it recognized the participants. They played the same game, but the sequence of rounds, and the questions Misty asked were slightly adjusted to reduce boredom. After the game, Misty used a similar social control strategy as in Session 1, with slight variations (Fig. 1). After the break, participants completed the same relationship questionnaire as session 1. Thereafter, the experimenter escorted participants to fill out a final questionnaire asking them to evaluate the strategies and provide demographic information. Participants were then debriefed, and compensated, and the experiment concluded.

F. *Statistical Analysis*

Descriptive statistics were used to provide an initial overview of the data. For the interim analysis, linear mixed-effects models were used to examine the effects of strategy and session on behavioral measures. Paired *t*-tests were used to assess relationship change over sessions. For the final analysis, Wilcoxon rank-sum tests were used to compare differences in active time between the two strategies. Chi-squared tests were conducted to examine the association between strategy and the likelihood of standing up and staying active throughout. Linear regression models assessed the interaction between relationship and strategy on active time. An alpha level of 0.0294 [56] was used for all statistical tests. All data and analysis scripts (conducted using *R* 4.2.3) are available at https://osf.io/hv9q5/.

IV. RESULTS

A. *Interim Data Analysis*

*1) Descriptive Analysis.* In session 1, participants receiving the relationship-focused strategy (*n* = 33) spent, on average, 223s (*SD* = 100s) being active, whereas those receiving the target-focused strategy (*n* = 33) spent 174s (*SD* = 116s) on average. In session 2, those receiving the relationship-focused strategy were active for an average of 165s (*SD* = 128s), while those receiving the target-focused strategy averaged 115s (*SD* = 119s). The histogram (Fig. 4, left) illustrates the active time distribution across both sessions for each strategy. In the relationship-focused condition, 15 participants remained active throughout the entire break (active time ≥ 300s) in session 1, and this number decreased to 8 in session 2. One participant did not stand up at all during session 1, and this number rose to 6 in session 2. In the target-focused condition, 8 participants stayed active throughout the break in session 1, while 4 did not stand up at all. By session 2, only 3 participants remained active for the entire break, and 12 did not stand up at all.

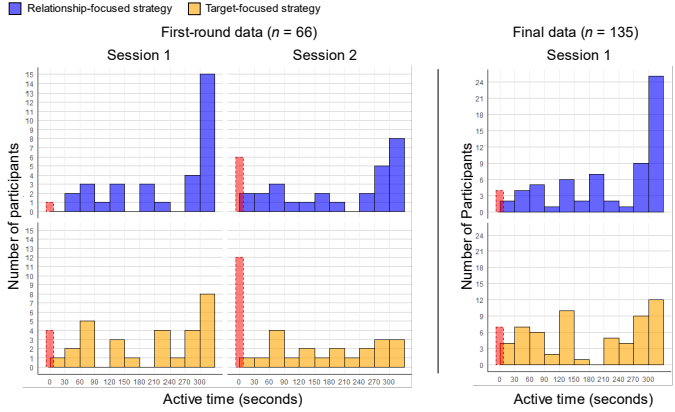

Fig. 4. Distribution of active time across two sessions for each strategy. Bars indicate the number of participants within 30-second left-closed intervals (i.e., [a, b)). The red bar denotes those who did not stand up at all (active time = 0).

*2) Effects of Strategy on Behavior.* Participants' active time was analyzed using a linear mixed-effects model with strategy, session, and their interaction as fixed effects, and participant ID as a random effect to account for individual variability. The target-focused strategy and session 1 were set as reference categories. The analysis revealed a significant effect of session, with session 2 leading to a significantly reduced active time ($\beta = -59.09$, $p = .001$). There was no significant main effect of strategy ($\beta = 48.70$, $p = .09$) or interaction between strategy and session ($\beta = 1.42$, $p = .95$).

The likelihood of participants standing up was analyzed using a generalized linear mixed model with a binomial family. The model included strategy and session as fixed effects, with a random intercept for participants. The results showed a significant main effect of the session, with session 2 leading to a substantial drop in the likelihood of standing up to session 1 ($\beta = -15.43$, $p < .001$). The strategy did not have a significant main effect ($\beta = 0.99$, $p = .75$). Another generalized linear mixed model examining the effect of strategy and session on the likelihood of staying active throughout found a similar pattern: session had a significant negative effect ($\beta = -15.45$, $p < .001$), but strategy did not ($\beta = 1.02$, $p = .73$).

Given the interim analysis of the first-round data, we did not find evidence that strategy impacts people's behaviors. However, considering the noticeable mean difference in the active time (49s) between the two strategies, we assumed that the effect of strategy might become evident with a larger sample size. Thus, the second round of data collection was launched.

*3) Relationship Strength over Sessions.* Pearson correlation analyses revealed significant positive correlations among the four relationship variables (all $r \geq 0.55$, all $p < .001$). One-tailed paired *t*-tests (Table II) showed that attachment, closeness, and companionship did not significantly increase from session 1 to session 2. Although the relationship stage (the sliding scale) showed a significant increase, the mean difference ($\Delta M = 4.71$) was very small, representing less than 5% of the 0-100 scale. Given these findings, it can be concluded that repeated interaction did not strengthen the relationship. Therefore, we decided to omit the second session in the second round of data collection.

TABLE II. ONE-TAILED PAIRED T-TESTS FOR RELATIONSHIP VARIABLES, WITH THE ALTERNATIVE HYPOTHESIS: SESSION 2 > SESSION 1.

|  | $M_{S1}$ | $SD_{S1}$ | $M_{S2}$ | $SD_{S2}$ | $\Delta M$ | $t$ | $p$ (one-tailed) |
|---|---|---|---|---|---|---|---|
| Attachment | 4.33 | 1.24 | 4.04 | 1.23 | −0.29 | −3.26 | .99 |
| Closeness | 4.65 | 1.28 | 4.61 | 1.28 | −0.04 | −0.49 | .69 |
| Companionship | 8.24 | 1.36 | 8.22 | 1.34 | −0.02 | −0.18 | .57 |
| Relationship Stage (Slider) | 48.00 | 24.76 | 52.71 | 25.72 | 4.71 | 2.91 | .002 |

*B. Final Data Analysis*

*1) Descriptive Analysis.* Only data from session 1 were included in the final data analysis. Participants receiving the relationship-focused strategy ($n = 68$) spent on average 207s ($SD = 106$s) being active, while those receiving the target-focused strategy ($n = 67$) spent 164s ($SD = 114$s) being active. For the relationship-focused strategy, 25 participants remained active throughout the entire break, and 4 participants did not stand up at all. For the target-focused strategy, there was a wider variation, with 12 participants remaining active throughout the entire duration and 7 did not stand up at all (Fig. 4, right).

*2) Effects of Strategy on Behavior.* A Wilcoxon rank-sum test examined the effect of strategy on active time. The test revealed that the relationship-focused strategy led to significantly longer active time than the target-focused strategy ($W = 2843$, $p = .01$, $r = 0.21$). Pearson's chi-squared tests revealed a significant association between strategy and participants' likelihood of staying active throughout ($\chi^2 = 5.12$, $p = .02$), but no significant association between strategy and the likelihood of standing up ($\chi^2 = 0.43$, $p = .51$).

*3) Effect of Relationship on Behavior.* Four separate linear regression models were conducted to examine the effect of human-robot relationships on active time. In each model, strategy and one relationship variable were included as predictors, along with their interaction terms, using the target-focused strategy as the reference category. The first model, examining the effects of strategy and attachment, revealed significant main effects of attachment ($\beta = 34.74$, $p = .002$) and strategy ($\beta = 205.08$, $p = .002$). A significant interaction between strategy and attachment was observed ($\beta = -36.33$, $p = .01$). These results indicate that higher attachment was associated with greater active time in the target-focused group, while the effect of attachment was weaker in the relationship-focused group. The second model, which examined the effects of strategy and relationship stage, similarly revealed significant main effects for the relationship stage ($\beta = 1.58$, $p = .004$) and strategy ($\beta = 147.56$, $p < .001$). Their interaction was also significant ($\beta = -2.08$, $p = .006$). The model with strategy and closeness showed a significant main effect of strategy ($\beta = 167.83$, $p = .02$), but not for closeness ($\beta = 18.47$, $p = .10$) or their interaction ($\beta = -26.63$, $p = .08$). Finally, the model with strategy and companionship did not reveal any significant main or interaction effects. Interaction plots (Fig. 5) visualized that higher attachment and relationship stages were associated with increased active time in the target-focused condition, but not in the relationship-focused condition.

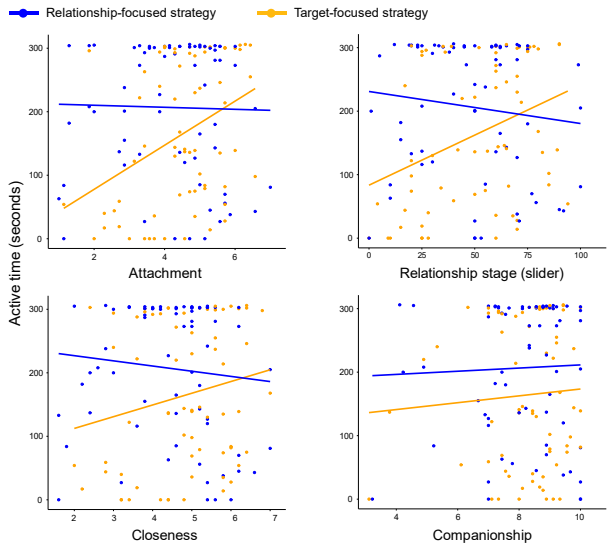

Fig. 5. Interaction between strategy and relationship variables on active time.

*4) Cognitive Evaluation of Strategy.* The average level of social appropriateness for both strategies can be interpreted as high, exceeding the midpoint of the 5-point scale ($M_{relationship}$ = 4.27, $SD_{relationship}$ = 0.79; $M_{target}$ = 4.32, $SD_{target}$ = 0.72). The average level of perceived threat to freedom can be interpreted as low, falling below the midpoint of the 7-point scale ($M_{relationship}$ = 2.56, $SD_{relationship}$ = 1.30; $M_{target}$ = 2.43, $SD_{target}$ = 1.37). Wilcoxon rank-sum tests indicated no significant difference in appropriateness ($W$ = 2188, $p$ = .69) or perceived threat ($W$ = 2438.50, $p$ = .48) between the two strategies.

## V. DISCUSSION

This study examined whether relationship-focused or target-focused strategies initiated by a social robot may differ in their impact on people's behavioral reactions (RQ1), whether repeated interactions may strengthen the perception of human-robot relationships (RQ2), and how relationship perception may moderate the effectiveness of the strategies (RQ3).

### A. Relationship-Focused versus. Target-Focused Strategy

We found that although the type of strategy did not significantly impact whether participants would immediately stand up, it had a significant impact on their active time, resulting in a substantial mean difference. This suggests that relationship-focused strategies were more effective in encouraging active behaviors compared to target-focused strategies. This result aligns with interpersonal research, which highlights that leveraging relational incentives is often more powerful than using factual health information [17, 18]. Two theoretical perspectives may help explain this result. First, according to interdependence theory, people in close relationships tend to shift their motivation from self-interest to relationship-oriented goals [33, 43, 73]. In this study, the relationship-focused strategy used by the robot implied the existence of a close relationship. By expressing care for participants and a desire for their well-being, the robot positioned participants in a relational role, potentially evoking a sense of responsibility to fulfill the robot's wishes. Second, attachment theory suggests that humans possess an innate drive to care for those perceived as dependent on them [10]. The robot's expression like "It would make me happier if you..." explicitly conveyed its dependence. Combined with the robot's child-like features—such as its small body, large eyes, and cute voice—the relationship-focused strategy may have activated people's altruistic motivation, thereby influencing their behavior more strongly.

Regarding cognitive evaluations, no significant differences were observed between the two types of strategies—neither was perceived as particularly threatening or inappropriate. This contrasts with the previous video-based study [72], which suggested that relationship-focused strategies from robots heightened feelings of threat. Three factors might explain this discrepancy. First, the prior study employed a within-subjects design, exposing participants to multiple strategies and allowing for direct comparisons. This experimental design likely heightened participants' sensitivity to the potential threats conveyed in the strategies. Second, the previous study was conducted online, with no real human-robot interaction. In contrast, participants in this study experienced a game with a physically present robot. This authentic and meaningful interaction likely softened the perceived forcefulness or manipulative intent of the robot's influence strategies. Previous studies have also shown that physically present robots tend to be perceived more positively than virtual robots, even when both exhibit identical behaviors [45]. Third, in the previous online study, participants only imagined that they were being seated. In such a hypothetical context, the strategies might seem unfounded and evaluated purely based on their message components. In contrast, the current study placed participants in a scenario where they had been actually seated for a while, likely making the robot's persuasion feel more relevant and reasonable. Taken together, the results of the current study are arguably more representative of people's perceptions in real HRI contexts.

### B. Repeated Interaction Didn't Enhance the Relationship

Contrary to our expectations, repeated interaction did not strengthen human-robot relationships, as most relationship measures showed no significant changes from session 1 to session 2. One exception is the sliding scale of the relationship stage, which showed a small but significant increase. This may be due to the inherent greater sensitivity of the sliding scale compared to the 7-point Likert scale. With a wider range (1-100), participants can express more subtle shifts in their perceptions, whereas a 7-point scale may force them to simplify their feelings. However, given that the increase was very small (less than 5%), it does not provide sufficient evidence that repeated sessions had a meaningful impact.

This result indicates that forming relationships with robots involves more complex processes than simple repeated exposure. According to Knapp's Relational Development Model [36], relationship progression requires a series of increasingly deeper interaction patterns, moving from initiation to experimentation and intensification. In our study, while the first game session may have facilitated an initial impression of the robot, the repetitive nature of the second session may have hindered the relationship from advancing to deeper stages. This finding indicates that future research that aims at fostering closer human-robot relationships should not only consider the temporal aspect but also implement richer, more dynamic interaction strategies.

*C. The Impact of Relationship on the Impact of Strategy*

We found that participants' relationship with the robot—specifically attachment and perceived relationship stage—positively influenced their behaviors. In the target-focused group, those more attached to the robot exhibited longer active times. This finding aligns with prior research in interpersonal contexts, which suggests that relationship closeness enhances the success of persuasion efforts [24, 40, 61]. However, an unexpected result was that this effect was not observed with the relationship-focused strategies. One possible explanation is the limited observational period of this experiment. The relationship-focused strategies are inherently more powerful than target-focused ones, leading more participants to remain active during the break. However, because the break lasted only 5 minutes, there may not have been sufficient time to observe a gradual increase in active behaviors driven by deeper emotional perceptions. In summary, although the impact of relationship dynamics emerged only with the target-focused strategies, this finding provides evidence that such dynamics play a significant and non-negligible role in enhancing a robot's influence.

*D. The Impact of Session on Behavior*

Beyond our expectations, the repeated sessions had a significant impact on people's behavioral reactions. In session 2, fewer participants stood up immediately, their active time decreased, and fewer participants remained active throughout the break compared to session 1. One possible explanation for this compliance decline is the novelty effect. In the first session, the unfamiliarity and intrigue of interacting with the robot likely drove higher compliance, as participants were curious to see how the robot would respond or how the experiment would proceed. By session 2, however, participants might have realized that the robot would not provide feedback for their actions, leading to diminished interest and less active responses. Similar findings have been reported in [3], where participants' motivation decreased when a conversational agent repeatedly delivered the same advice, eventually resulting in boredom and annoyance. Another possible explanation is the natural decline in politeness that comes with increased social familiarity. According to politeness theory [7], people are typically more polite when interacting with strangers compared to familiar acquaintances. In this study, participants may have been more compliant in session 1 out of a social obligation to be polite. By session 2, as they became more familiar with the robot, this sense of obligation may have diminished, leading to reduced politeness and, consequently, lower compliance levels.

*E. Implication for HRI Research and Design*

This study provides several valuable implications for the HRI community. Theoretically, it bridges insights from interpersonal social control literature with the HRI domain, demonstrating that relationship-focused strategies can be effectively utilized by robots to influence human behavior. The observed impact of human-robot relationships on participants' active time also aligns with findings from interpersonal persuasion research, emphasizing that the quality of relationships serves as a significant contextual moderator in the persuasion process. Practically, these findings offer a heuristic for designers aiming to enhance robots' capacity to promote health behavior change, suggesting that robots' health communication strategies can move beyond traditional informational exchanges to appeal to relationship-oriented motivations. Methodologically, this study introduces a novel approach to measuring health behavior change in HRI research. As noted in a recent systematic review [48], most prior experimental works relied on highly artificial and abstract tasks (e.g., survival games [12], decision-making games [26, 28, 63]) to evaluate social robots' persuasiveness. In contrast, our experimental paradigm situates robotic persuasion in a more realistic context and offers new ways to observe and quantify more meaningful behavior change, which could represent a valuable step forward for the study of persuasive social robots.

*F. Limitation and Future Work*

This study has several limitations. First, the task was relatively short, requiring participants to remain seated for only 10 minutes. This limited sitting duration may have diminished the perceived necessity of the behavior change appeals, potentially constraining the impact of the designed strategies—particularly the target-focused strategies, as participants might not have felt a strong motivation to pursue health benefits. While our findings revealed differential impacts of the strategies, we recommend that future research extend the task duration to further validate these results and explore the effectiveness of these strategies in scenarios with more pronounced health implications. Second, it is important to note that this study focused on observing one-off health actions rather than genuine health behavior change, which typically requires sustained healthy actions over time. As a preliminary step toward the broader goal of designing social robots to promote health behavior change, this study concentrated on scrutinizing nuanced dimensions of health communication design, with long-term effectiveness evaluation beyond its scope. Moving forward, we are currently planning to conduct real-world, longitudinal HRI studies to better understand how these strategies can facilitate lasting health behavior change.

## VI. CONCLUSION

Designing social robots to influence health behaviors is an emerging research trend, with the search for strategies to boost compliance with recommended behaviors becoming a critical area of focus. This experiment explored the effects of two distinct types of social control strategies employed by a robot in a scenario where the robot persuaded participants to stand up and walk after a period of sitting. Our results revealed that relationship-focused strategies were more effective at motivating participants to stay active than traditional target-focused strategies. Moreover, participants' relationship with the robot positively moderated the impact of target-focused strategies, which highlights the importance of considering human-robot relationship dynamics when designing social robots as persuasive agents. Additionally, repeated interaction sessions did not significantly strengthen the human-robot relationship. This suggests that time alone may not suffice for human-robot relationship progression. Envisioning a future where social robots could evolve into relational partners for humans, this study provides a fresh perspective on health communication design for social robots and opens new discussions on how human-robot relationships can be shaped and leveraged to influence human behaviors.